\documentclass{PoS}

\usepackage{amsmath,subfigure,latexsym,amssymb,cite}

\newcommand{\be}{\begin{equation}}
\newcommand{\ee}{\end{equation}}
\newcommand{\nn}{\nonumber}
\newcommand{\bea}{\begin{eqnarray}}
\newcommand{\eea}{\end{eqnarray}} 

\newcommand{\la}{\langle}
\newcommand{\ra}{\rangle}

\newcommand{\R}{{\kern+.25em\sf{R}\kern-.78em\sf{I} \kern+.78em\kern-.25em}}
\newcommand{\RR}{{\kern+.25em\sf{R}\kern-.6em\sf{I} \kern+.6em\kern-.25em}}
\newcommand{\N}{{\kern+.25em\sf{N}\kern-.78em\sf{I} \kern+.78em\kern-.25em}}

\newcommand{\Z}{\mathbb{Z}}

\newcommand{\gtapprox}{\raisebox{-0.5ex}{$\,\stackrel{>}{\scriptstyle\sim}\,$}}
\newcommand{\ltapprox}{\raisebox{-0.5ex}{$\,\stackrel{<}{\scriptstyle\sim}\,$}}

\makeatletter
\@addtoreset{equation}{section}
\makeatother

\title{\vspace*{-1.2cm}
\begin{flushright}
\texttt{\footnotesize CERN-TH-2016-212}\\
\end{flushright}
\vspace*{0.5cm}
The Slab Method to Measure the Topological Susceptibility\thanks{We
thank Stephan D\"{u}rr, Massimo D'Elia and
Marc Wagner for helpful discussions. This work was supported 
by the Mexican {\it Consejo Nacional de Ciencia y 
Tecnolog\'{\i}a} (CONACYT) through projects CB-2010/155905 and 
CB-2013/222812, by DGAPA-UNAM, grant IN107915, and by the 
{\it Helmholtz International Center for FAIR} within the 
framework of the LOEWE program launched by the State of Hesse.
A.D.\ acknowledges support by the Emmy Noether Programme 
of the DFG (German Research Foundation), grant WA 3000/1-1. The 
computations were performed on the cluster of ICN/UNAM, and
on the LOEWE-CSC and FUCHS-CSC high-performance 
computer of Frankfurt University.}}

\ShortTitle{Slab Method}

\author{\speaker{Wolfgang Bietenholz}$^{\rm \,\, a}$,
Krzysztof Cichy$^{\rm \,\, b,c}$, Philippe de Forcrand$^{\rm \,\, d,e}$, \newline
Arthur Dromard$^{\rm \,\, b,f}$ and Urs Gerber$^{\rm \,\, a,g}$
\ \\
\vspace*{1mm}
\ \\
\ $^{\rm \ a}$ Instituto de Ciencias Nucleares, 
Universidad Nacional Aut\'{o}noma de M\'{e}xico\\
~~~~A.P.\ 70-543, C.P.\ 04510 Ciudad de M\'{e}xico, 
Mexico \vspace*{1mm} \\
\ $^{\rm \, b}$ Goethe-Universit\"{a}t Frankfurt am Main,
Institut f\"{u}r Theoretische Physik \\
~~~~Max-von-Laue-Stra\ss e 1, D-60438 Frankfurt am Main, Germany
\vspace{1mm} \\
\ $^{\rm \, c}$ Faculty of Physics, Adam Mickiewicz University,
Umultowska 85, 61-614 Poznan, Poland \vspace*{1mm} \\
\ $^{\rm \, d}$ Institut f\"{u}r Theoretische Physik, ETH Z\"{u}rich,
CH-8093 Z\"{u}rich, Switzerland \vspace*{1mm} \\
\ $^{\rm \, e}$ CERN, Theory Division,
CH-1211 Geneva 23, Switzerland \vspace*{1mm} \\
\ $^{\rm \, f}$ Institut f\"{u}r Theoretische Physik, Universit\"{a}t 
Regensburg, D-93040 Regensburg, Germany \vspace*{1mm} \\
$^{\rm \, g}$ \ Instituto de F\'{\i}sica y Matem\'{a}ticas,
Universidad Michoacana de San Nicol\'{a}s de Hidalgo\\
~~~~Edificio C-3, Apdo.\ Postal 2-82, C.P.\ 58040, 
Morelia, Michoac\'{a}n, Mexico \vspace*{3mm}

E-mail: \email{wolbi@nucleares.unam.mx} \\ }

\abstract{In simulations of a model with topological sectors,
algorithms which proceed in small update steps
tend to get stuck in one sector, especially on fine 
lattices. This distorts the numerical results; in 
particular it is not straightforward to measure the 
topological susceptibility $\chi_{\rm t}$. We test 
a method to measure $\chi_{\rm t}$ even if configurations 
from only one sector are available. It is based on the topological
charges in sub-volumes, which we denote as ``slabs''.
This enables the evaluation of $\chi_{\rm t}$, as we demonstrate 
with numerical results for non-linear $\sigma$-models and for
2-flavour QCD. In the latter case, the gradient flow is applied 
for the smoothing of the gauge configurations, and the slab method 
results for $\chi_{\rm t}$ are stable over a broad range of flow times.}

\FullConference{34th International Symposium on Lattice Field Theory\\
		24-30 July 2016\\
		University of Southampton, UK}

\begin{document}

\section{The topological susceptibility $\chi_{\rm t}$}
\vspace*{-3mm}

In models with topological sectors,
a quantity of interest is the topological susceptibility
\be
\vspace*{-1.5mm}
\chi_{\rm t} = \frac{1}{V} \left( \la Q^{2} \ra - \la Q \ra^{2} \right) \ ,
\qquad Q ~:~ {\rm topological~charge}, \quad 
V ~:~ {\rm volume.}
\ee
We are going to address settings with parity invariance, 
where $\chi_{\rm t}$ simplifies due to $\la Q \ra = 0$. 

A prominent application is the Witten-Veneziano formula,
as a quantitative solution to the U(1) problem:
for three massless quark flavours and large $N_{\rm c}$, the
$1/N_{\rm c}$ corrections yield 
$\chi_{\rm t}^{\rm quenched} \simeq F_{\pi}^{2} M_{\eta '}^{2}/6$,
where $F_{\pi}^{2} \propto N_{\rm c}$, and $M_{\eta '}^{2} \propto 1/N_{\rm c}$. 
For QCD with dynamical quarks, there is a similar relation 
to a putative axion mass and decay constant,
$\chi_{\rm t} \simeq F_{\rm axion}^{2} M_{\rm axion}^{2}$. Hence the value 
of $\chi_{\rm t}$ (at finite temperature) is relevant for the question 
whether or not the axion is a valid Cold Dark Matter candidate; for a 
review and recent lattice results, see {\it e.g.}\ Refs.\ \cite{axion}.

$\chi_{\rm t}$ can only be determined non-perturbatively, hence 
numerical measurements in lattice simulations are appropriate.
If a Monte Carlo history changes the topological sector 
frequently, it is straightforward to measure $\la Q^{2} \ra$
(once one has defined the topological charge of the lattice
configurations). This is the case for instance in quenched
QCD, simulated with the heatbath algorithm at lattice 
spacing $a > 0.1 \, {\rm fm}$;
an example is described in Ref.\ \cite{Stani}.

Another direct approach is to measure (in lattice units)
$\chi_{\rm t} = \sum_{x \in V} \la q_{0} q_{x} \ra$,
where $q_{x}$ is the topological charge density. This has been 
applied successfully to $2+1$ flavour QCD \cite{MILC}. The 
long-distance correlation function was fitted to an expected linear 
combination of modified Bessel functions $K_{1}$, where the 
phenomenological values of $M_{\eta}$ and $M_{\eta'}$ were inserted.

As we decrease $a$, however, the topological sectors
are separated by higher and higher potential barriers.
Then an algorithm which performs small
update steps tends to get stuck in one topological sector for
a very long (computation) time. 
According to Ref.\ \cite{alphascal}, the autocorrelation time 
with respect to $Q$, $\tau_{Q}$, in simulations of SU($N$) Yang-Mills
theories (with the Wilson lattice action, and alternating
overrelaxation and heatbath steps), is compatible with an exponential 
growth, or a high power, in $1/a$. For QCD, Ref.\ \cite{Rainer}
observed a behaviour $\propto (1/a)^{z}$ with $z\simeq 5$ in the
quenched case, and similar with dynamical quarks, represented by 
$O(a)$-improved Wilson fermions (though $z$ is less accurate).
Dynamical chiral quarks make the growth of $\tau_{Q}$ even worse. 

One way to deal with this issue is to modify the algorithm such
that changes of $Q$ become more frequent; such efforts are
reviewed in Ref.\ \cite{Endres}. A different approach suggests 
the use of open boundary conditions in Euclidean time \cite{openbc}, 
which removes the topological sectors, $Q \in \R$, but it breaks lattice
translation invariance. Here we address yet another concept, 
which aims at determining $\chi_{\rm t}$ even
from data in one fixed (``frozen'') topological sector.

One approach which --- in principle --- could be used for this
purpose is an approximation for some expectation value
$\la {\cal O} \ra$, if only measurements in fixed sectors, 
$\la {\cal O} \ra\vert_{Q}$, are available \cite{BCNW},
\vspace*{-2mm}
\be
\la {\cal O} \ra \vert_{Q} \simeq \la {\cal O} \ra
+ \frac{\rm const.}{V \chi_{\rm t}} \Big( 1 - \frac{Q^{2}}
{V \chi_{\rm t}} \Big) \ .
\vspace*{-1mm}
\ee
This is the beginning of an expansion in 
$1/(V \chi_{\rm t}) = 1/\la Q^{2}\ra$, 
extensions are discussed in Refs.\ \cite{MarcArt,BCNWnum}.
Once we have a set of results for $\la {\cal O} \ra\vert_{Q}$,
in different $V$ and $|Q|$, a fit provides values for the 
unknown (intensive) quantities: 
$\la {\cal O} \ra$, $\chi_{\rm t}$ and the const. 
A detailed numerical study \cite{BCNWnum}, in
a variety of models, shows that this works quite well for the 
determination of $\la {\cal O} \ra$ if suitable conditions are
fulfilled\footnote{Typically $\la Q^{2}\ra > 1$ is required,
and one should only involve sectors with small $|Q|$.\label{fn1}},
but the results for $\chi_{\rm t}$ are plagued by large uncertainties.

More successful for the determination of $\chi_{\rm t}$ --- though
exclusively devoted to that purpose --- is an approximation derived
in Ref.\ \cite{AFHO} (in a way similar to Ref.\ \cite{BCNW}),
\be  \label{AFHOeq}
\la q_{0} \, q_{x}\ra \vert_{|Q|, \, {\rm large}\, |x|} \simeq
- \frac{\chi_{\rm t}}{V} \Big( 1 - \frac{Q^{2}}
{V \chi_{\rm t}} \Big) \ .
\ee
One measures the left-hand side and searches for a plateau 
of the correlation function over long distances. This determines 
$\chi_{\rm t}$, under conditions similar to footnote \ref{fn1}. 
The problem is to resolve
tiny plateau values as the volume increases, but their suppression
can be compensated by computing all-to-all correlations \cite{qqpap}.

Here we discuss yet another, particularly simple approach, which 
we denote as the ``slab method''.

\vspace*{-3mm}
\section{The slab method}
\vspace*{-2mm}

The idea of the slab method was first mentioned in Ref.\ \cite{Phil}
and recently tested in the framework of $\sigma$-models \cite{slabpap}
and in two flavour QCD \cite{slabQCD}.
There is some similarity with the method of Ref.\ \cite{LSD}, and
with an instanton-liquid consideration in Ref.\ \cite{Jac}.

\begin{figure}
\vspace*{-2mm}
\begin{center}
\includegraphics[angle=0,width=.35\linewidth]{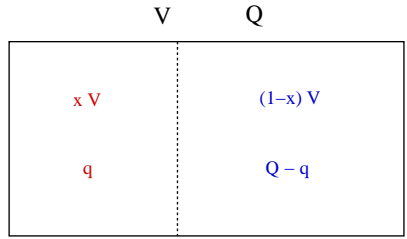}
\end{center}
\vspace*{-4mm}
\caption{Division of a volume $V$ into
{\em slabs} of sizes $xV$ and $(1-x)V$, with topological charges
$q$, $Q-q \in \R$.}
\vspace*{-4mm}
\label{slabfig}
\end{figure}

We assume a Gaussian distribution of the topological charge,
$
p(Q) \propto e^{- Q^{2} /(2 \chi_{\rm t} V)} 
$,
which is approximately confirmed, see below.
Next we split the volume $V$ into sub-volumes of sizes $xV$ and
$(1-x)V$ \ ($0 < x < 1$) --- which we denote as {\em slabs} --- 
as illustrated in Fig.\ \ref{slabfig}. For a configuration with
total topological charge $Q$, the slabs carry charges $q$ and
$Q-q$ (obtained by summing up the density).
Note that $q$ and $Q-q$ do not need to be integers, because the
face between the slabs is a non-periodic boundary.
At fixed $V$, $x$ and $Q$, the probabilities $p_{1}$, $p_{2}$ 
for the slab charges obey
\be  \label{slabeq}
p_{1}(q) \, \cdot \, 
p_{2}(Q-q) \, \propto \, 
\exp \Big( - \frac{q^{2}}{2 \chi_{\rm t} V x}\Big) \, \cdot \,
\exp \Big( - \frac{(Q-q)^{2}}{2 \chi_{\rm t} V (1-x)} \Big)
\, \propto \, \exp \Big( - \frac{1}{2  \chi_{\rm t} V} \
\frac{q{'}^{\, 2}}{x(1-x)} \Big) \ ,
\ee
where $q' := q - xQ$, and from $\la q \ra = xQ$ we infer
$\la q{'}^{\, 2} \ra = \la q^{\, 2} \ra - x^{2} Q^{2}$.
The idea is to measure $\la q^{2}\ra$ (and $\la q{'}^{\, 2} \ra$)
at various $x$. A fit of the $x$-dependence to the expected
parabola yields a value for $\chi_{\rm t}$.

\vspace*{-2mm}
\section{Results}
\vspace*{-2mm}

\subsection{Quantum rotor}

\begin{figure}
\vspace*{-2.5mm}
\begin{center}
\hspace*{-5mm}
\includegraphics[angle=270,width=.36\linewidth]{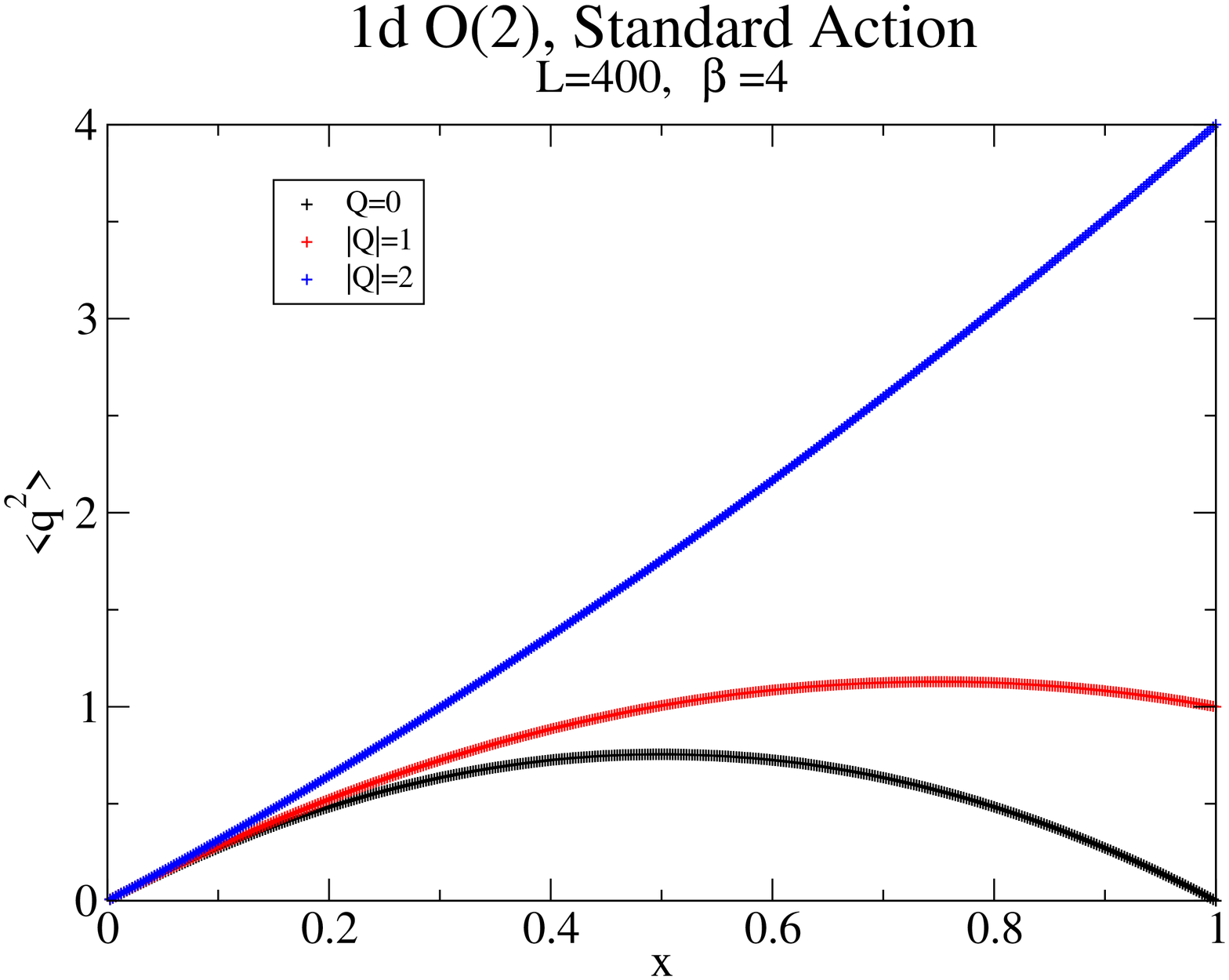}
\hspace*{-7mm}
\includegraphics[angle=270,width=.36\linewidth]{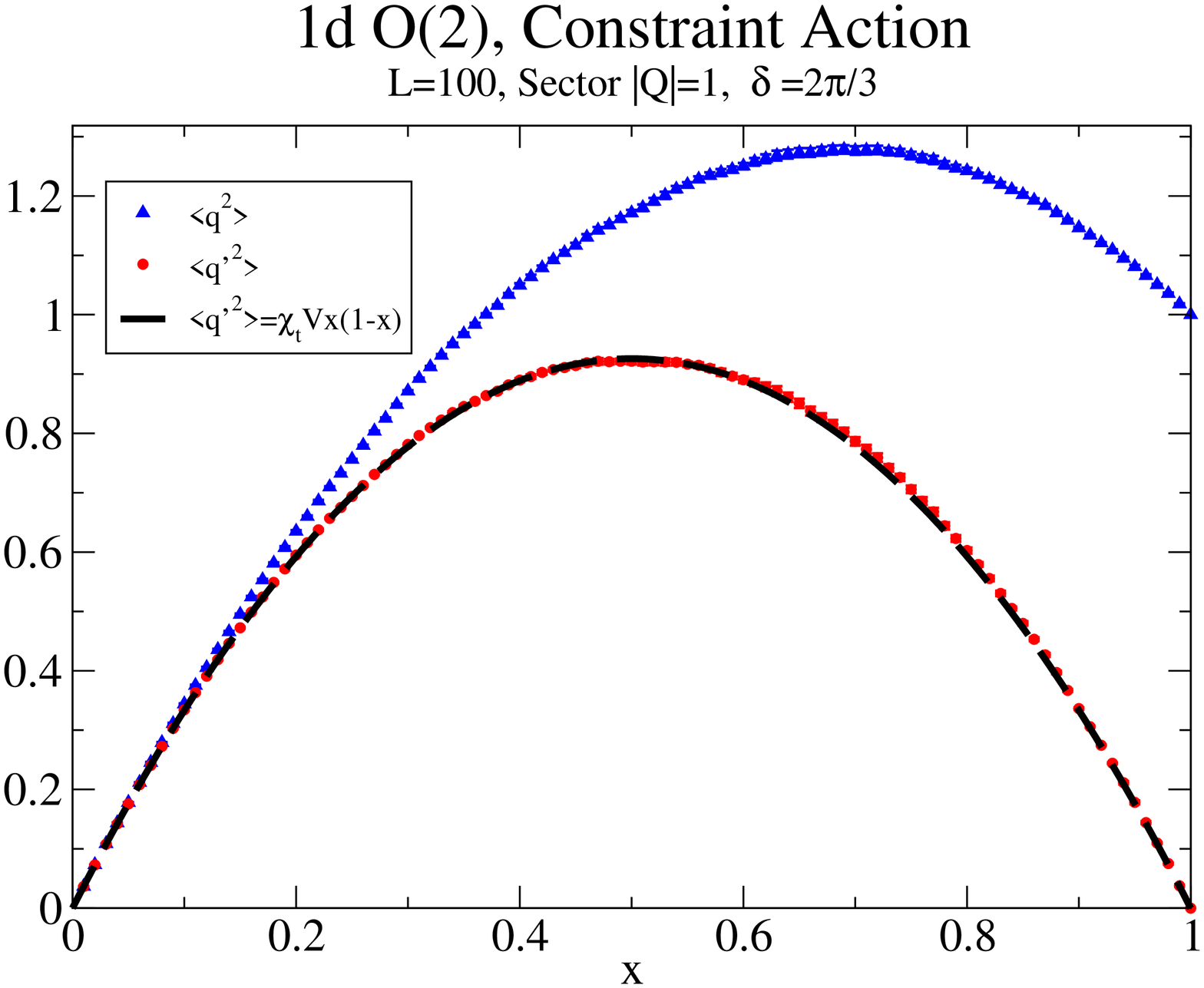}
\hspace*{-7mm}
\includegraphics[angle=270,width=.36\linewidth]{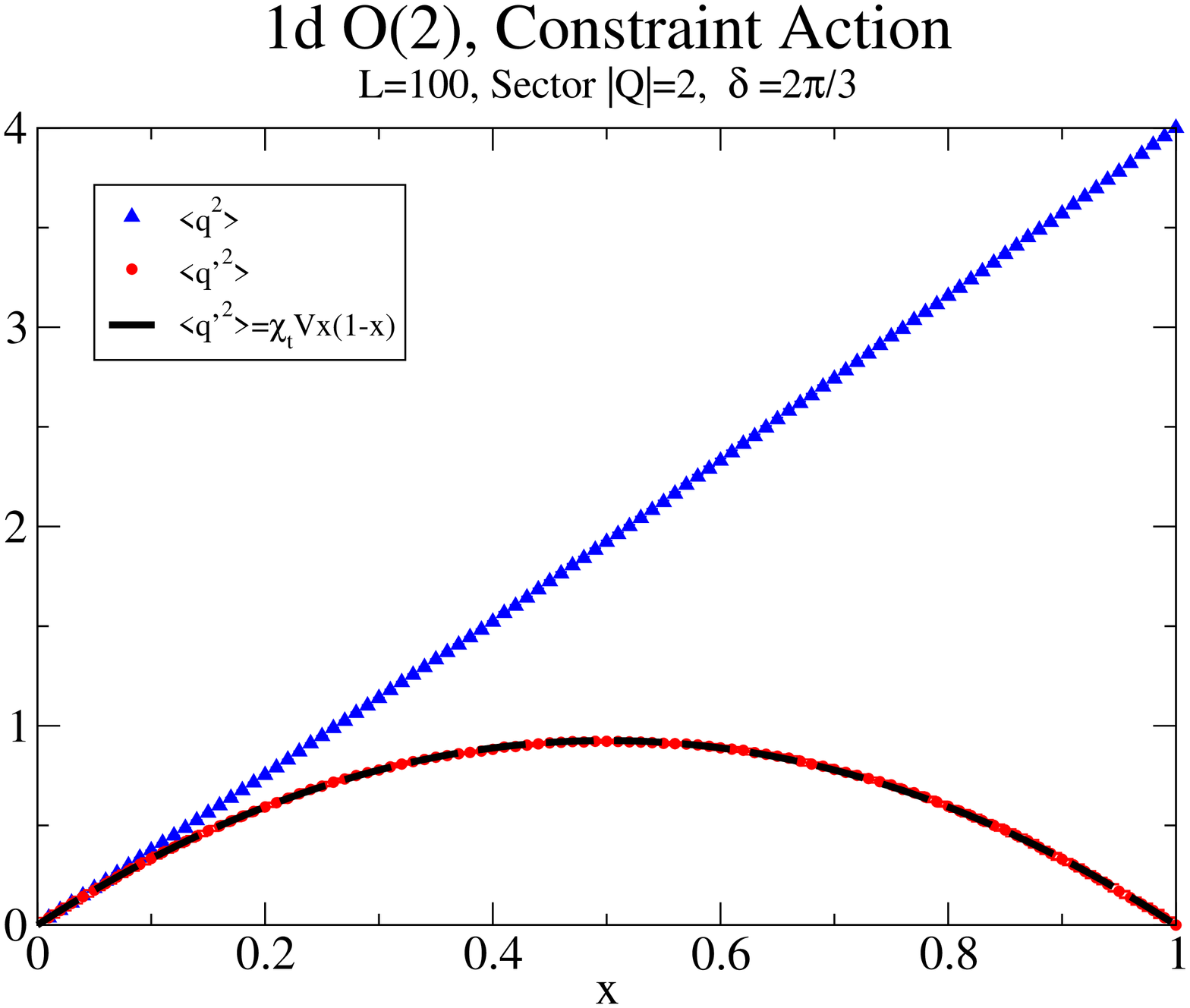}
\end{center}
\vspace*{-7.5mm}
\caption{Left: $\la q^{2}\ra$ measured for $S_{\rm standard} \,$ 
($L=400, \ \beta=4$) in the sectors $|Q|=0,\, 1,\, 2$.
Centre/right: $\la q^{2}\ra$ and $\la q{'}^{\, 2}\ra$ for 
$S_{\rm constraint} \,$
($L=100, \ \delta = 2 \pi /3$) at $|Q|= 1$ 
(centre) and $|Q|= 2$ (right).}
\label{parafit}
\vspace*{-1mm}
\end{figure}

We start with high-precision results for the
quantum rotor (or 1d XY model, or 1d O(2) model) \cite{slabpap}. 
Each site of a periodic lattice in Euclidean time carries an angular 
variable $\phi_{t}$, and we define the topological charge geometrically,
\be
Q [\phi ] = \frac{1}{2\pi} \sum_{t} \Delta \phi_t \ , \quad
\Delta \phi_t = (\phi_{t+1} - \phi_t ) \ {\rm mod} \ 2 \pi 
\in (-\pi, \pi] \ .
\ee
We consider three lattice actions,
\be
S_{\rm standard} [\phi ] = \beta \sum_{t} (1 - \cos (\Delta \phi_t) ) , \
S_{\rm Manton} [\phi ] = \frac{\beta}{2} \sum_{t} (\Delta \phi_t)^{2} , \
S_{\rm constraint} [\phi ] = \left\{ 
\begin{array}{ccc} 0 && | \Delta \phi_t | < \delta \ \ \forall t \\
+ \infty && {\rm otherwise.} \end{array} \right. \nn
\ee
Typical results for $\la q^{2}\ra$
and  $\la q{'}^{\, 2}\ra$ are shown in Fig.\ \ref{parafit}.
In each case they match the expected parabola to high
accuracy; this parabola connects $\la q^{2}\ra|_{x=0}=0$ with
$\la q^{2}\ra|_{x=1}=Q^{2}$, and $\la q{'}^{\, 2}\ra|_{x=0} =0$
with $\la q{'}^{\, 2}\ra|_{x=1} = 0$; the latter is predicted as
$L \, \chi_{\rm t} \, x (1-x)$.

\begin{figure}
\vspace*{-2mm}
\begin{center}
\includegraphics[angle=270,width=.43\linewidth]{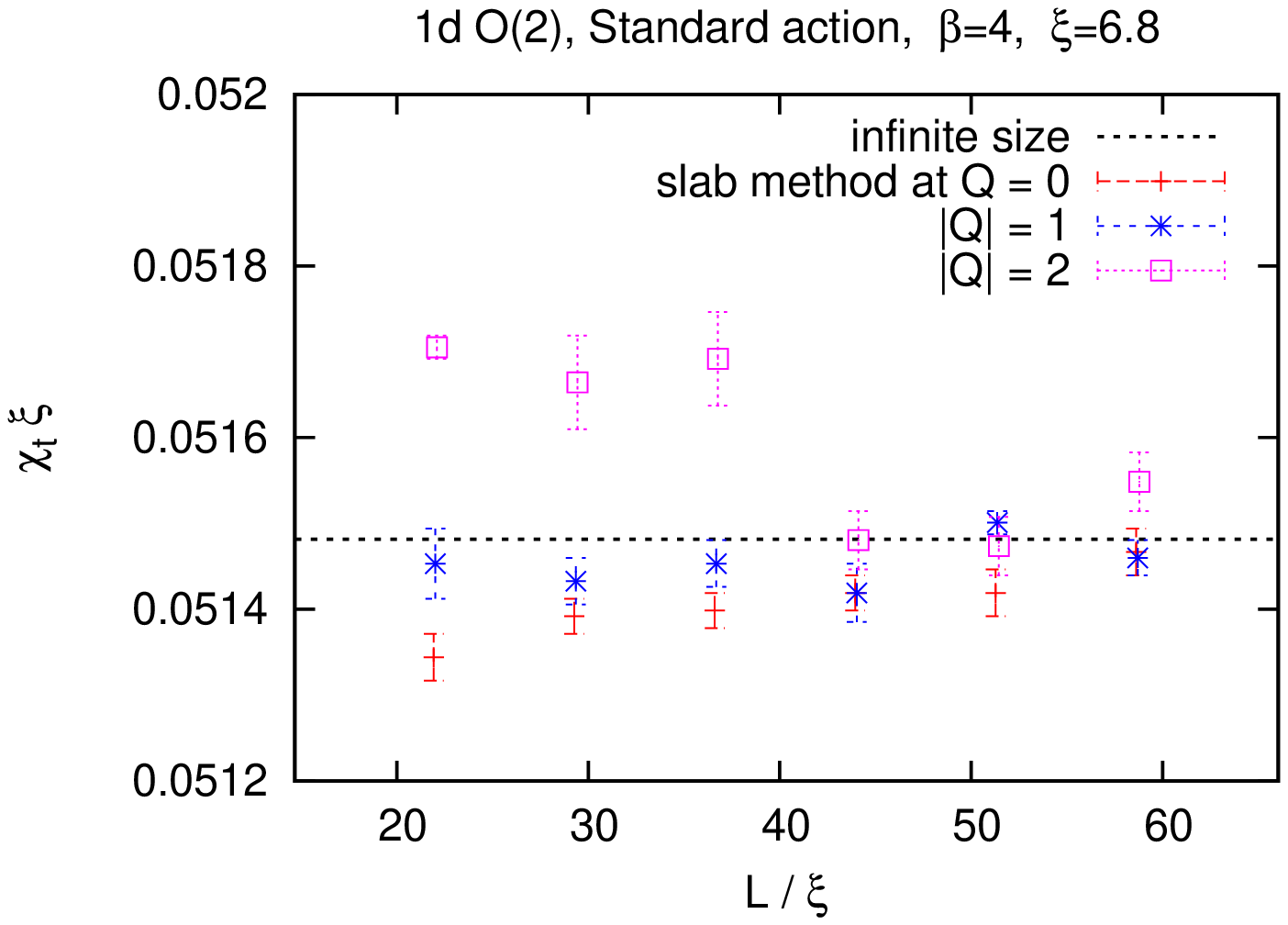}
\includegraphics[angle=270,width=.43\linewidth]{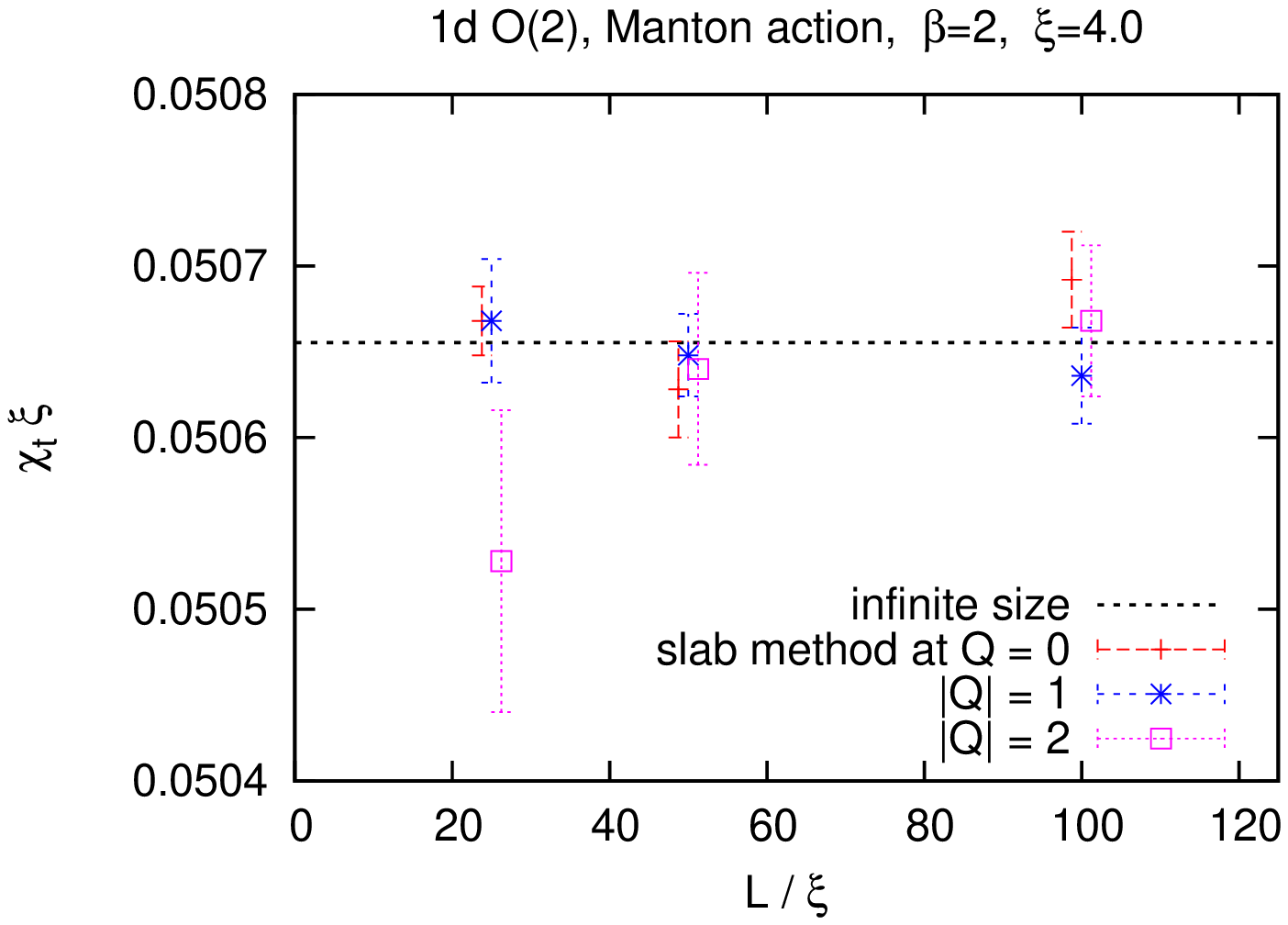}
\end{center}
\vspace*{-4mm}
\caption{The finite size scaling of $\chi_{\rm t} \, \xi$,
for the standard action at $\beta =4$, and for the Manton action
at $\beta =2$.}
\label{scaleplots}
\vspace*{-4mm}
\end{figure}

Now we consider the results for the scaling quantity
$\chi_{\rm t} \, \xi$, where $\xi$ is the correlation length.
For all three lattice actions the value is known analytically 
\cite{rot97,topact} in the thermodynamic limit, $L \to \infty$.
The plots in Fig.\ \ref{scaleplots} illustrate the convergence
towards these values (horizontal lines) at fixed $\beta$, for
increasing size. This convergence is manifest, but slow: 
in particular for the standard action there are permille
level finite size effects even at $L/\xi > 30$; these effects 
are enhanced for increasing $|Q|$.

\subsection{Heisenberg model}

\begin{figure}
\vspace*{-5mm}
\begin{center}
\includegraphics[angle=270,width=.454\linewidth]{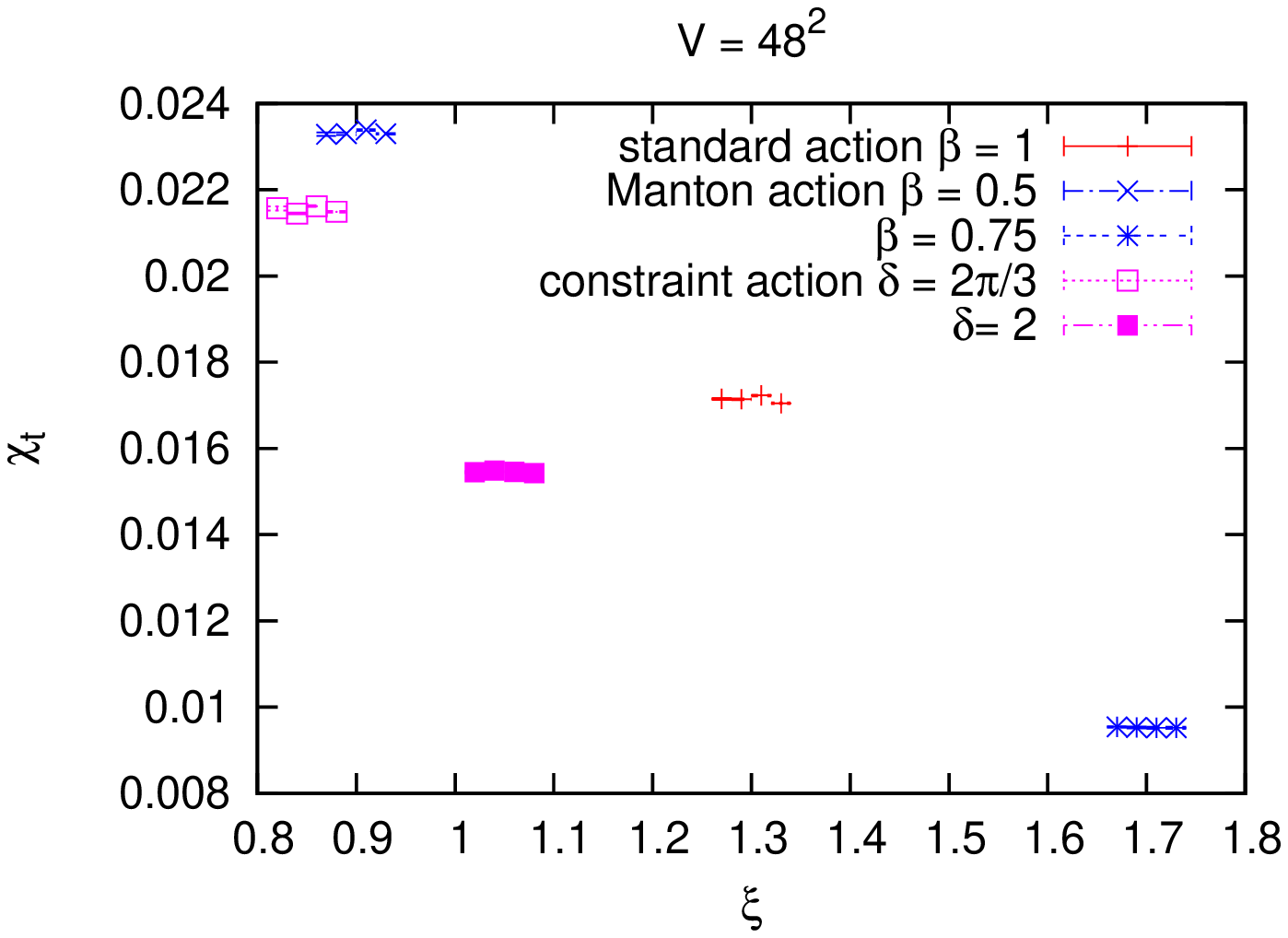}
\includegraphics[angle=270,width=.454\linewidth]{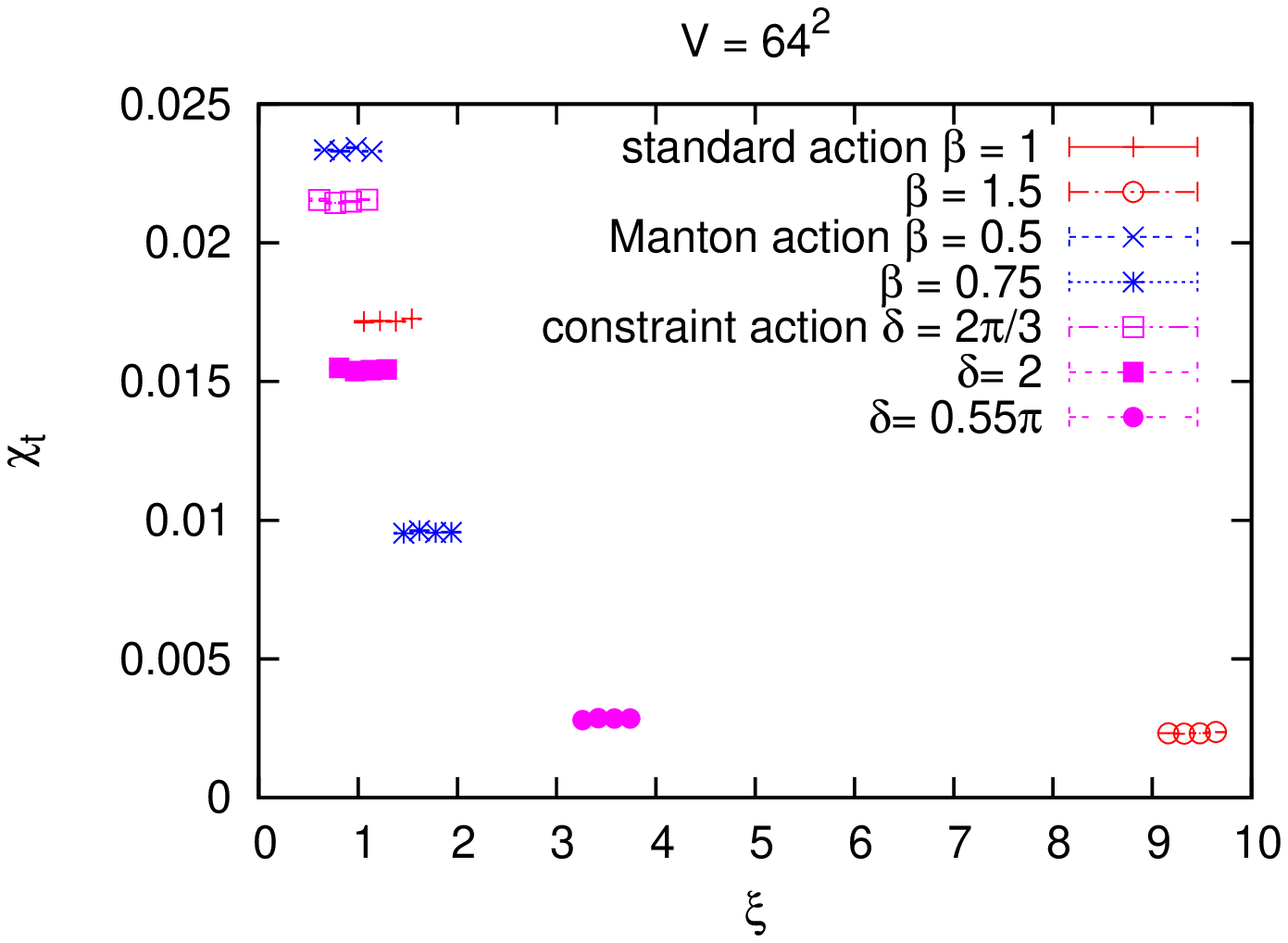}
\end{center}
\vspace*{-4mm}
\caption{Data for the 2d O(3) model in $V=48^{2}$ 
and $64^{2}$: each quadruplet of points shows 
(from left to right) the directly
measured $\chi_{\rm t}$, and the values obtained by the
slab method in the sectors $|Q|=0,\, 1,\, 2$.}
\vspace*{-3mm}
\label{O3res} 
\end{figure}

We proceed to the 2d Heisenberg model, or O(3) model.
Here the ``scaling term'', $\chi_{\rm t} \, \xi^{2}$,
diverges logarithmically in the continuum limit,
see {\it e.g.}\ Ref.\ \cite{topact}.
Hence we consider just $\chi_{\rm t}$ at finite $\xi$ (in
lattice units). Again we apply the geometric definition for $Q$
\cite{BergLuscher}, and we consider the three lattice actions, 
which are analogous to Subsection 3.1.
Fig.\ \ref{O3res} shows that the results are
very close to the directly measured values of $\chi_{\rm t}$;
those are precise in this case, thanks to the use of the Wolff 
cluster algorithm, which avoids topological freezing.
The data are given in Ref.\ \cite{slabpap}.

\begin{figure}
\begin{center}
\vspace*{-4mm}
\includegraphics[angle=270,width=.45\linewidth]{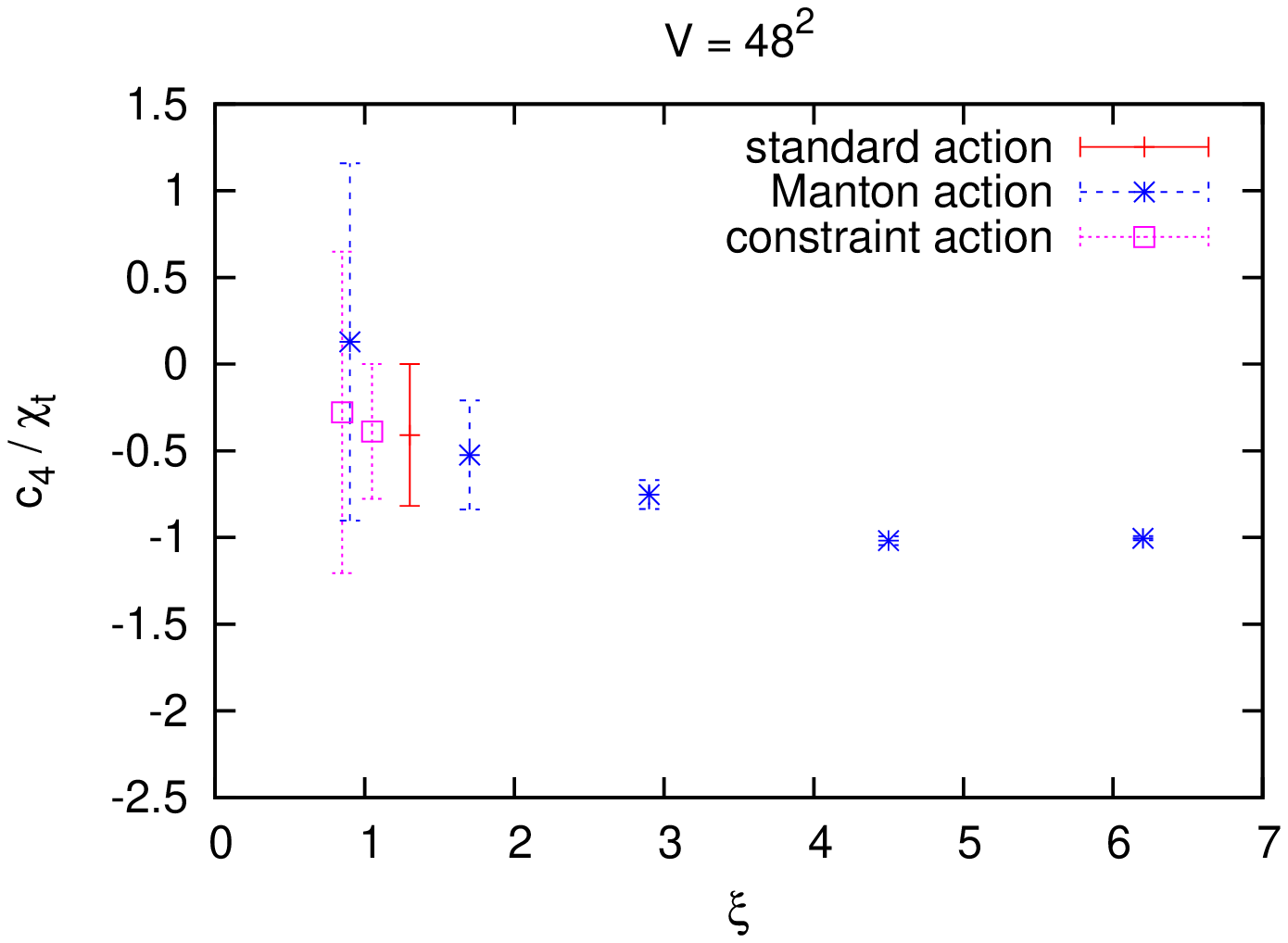}
\includegraphics[angle=270,width=.45\linewidth]{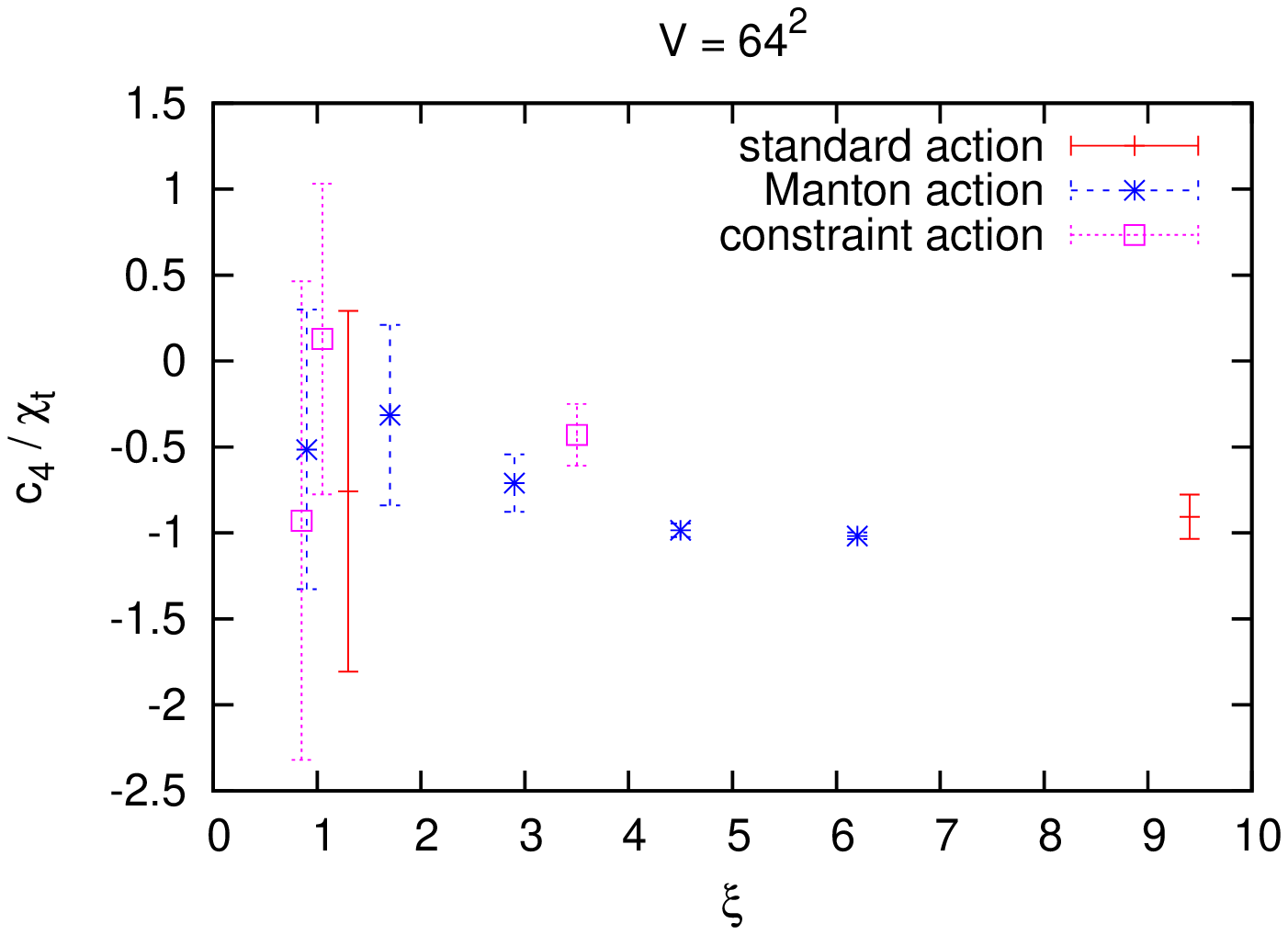}
\end{center}
\vspace*{-4mm}
\caption{Dependence of $c_{4}/\chi_{\rm t}$ on the correlation 
length $\xi$, for different lattice actions and volumes.}
\label{kurt}
\vspace*{-4mm}
\end{figure}

We also consider the kurtosis $c_{4}$,
\be
c_{4} = \frac{1}{V} \Big( 3 \la Q^{2} \ra^{2} - \la Q^{4} \ra \Big) \ ,
\ee
which represents a measure of the deviation from
a Gaussian distribution (where $c_{4}$ vanishes).
Fig.\ \ref{kurt} shows the convergence of the (dimensionless) ratio 
$c_{4}/\chi_{\rm t}$ in the continuum limit 
towards $\simeq -1$, the value for a dilute instanton gas;
this is best visible for the Manton action.\footnote{In $d=1$ the 
Manton action is classically perfect \cite{rot97}, which explains 
its excellent scaling behaviour. Apparently its 2d version was used 
first in Ref.\ \cite{slabpap}, and it has favourable properties as well.}
Comparing the two plots in Fig.\ \ref{kurt} suggests that --- in this 
regime --- the volume hardly affects the ratio 
$c_{4}/\chi_{\rm t}$.\footnote{This quantity has been investigated 
extensively in 4d SU(3) Yang-Mills theory, see {\it e.g.}\ Refs.\ 
\cite{Durr}. According to the latest studies, $c_{4}/\chi_{\rm t}$ 
converges of to a small but finite value around $-0.26$ 
in the continuum and infinite volume limit.}

\subsection{2-flavour QCD}

Finally we proceed to 2-flavour QCD, formulated with the
Wilson gauge action. The topological charge density is constructed
from the standard lattice field strength tensor. After smoothing,
$\sum_{x} q_{x}$ is slightly re-scaled (for optimal proximity
to integers \cite{alphascal}) and then rounded to $Q\in \Z$.

For the quarks we used twisted mass fermions (full twist,
bare mass $0.015$), which leads to a somewhat heavy pion mass,
$M_{\pi} \simeq 650 \, {\rm MeV}$ (here we are only interested in 
testing the slab method). The statistics involved $20 \, 000$
configurations, in $V = 16^{3} \times 32$ (and slab volumes 
$16^{3} \times 32 x$ and $16^{3} \times 32(1-x)$) 
at $\beta =3.9$, which implies a lattice spacing of 
$a \simeq 0.079 \, {\rm fm}$.

Smoothing was performed by the gradient flow (or Wilson flow 
in this case), with Runge-Kutta integration in the flow time $t$ 
(step sizes 0.01 and 0.001 yield consistent results).
The reference point proposed by L\"{u}scher \cite{Wilsonflow}, 
$t_{0}^{2} \la E \ra_{\rm plaquette} = 0.3$, requires the 
flow time $t_{0} = 2.42$.

\begin{figure}
\vspace*{-2mm}
\begin{center}
\includegraphics[angle=270,width=.45\linewidth]{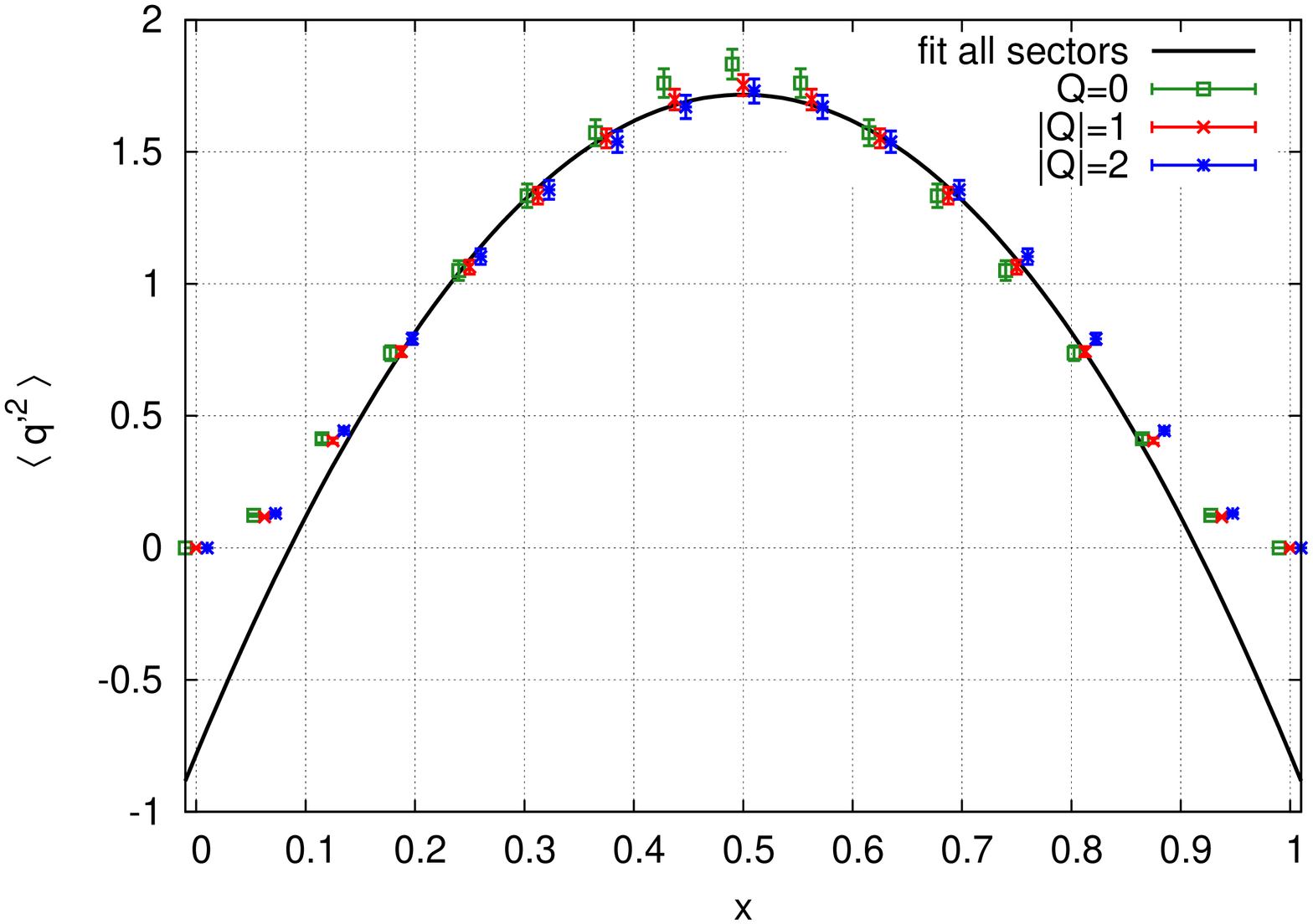}
\includegraphics[angle=270,width=.45\linewidth]{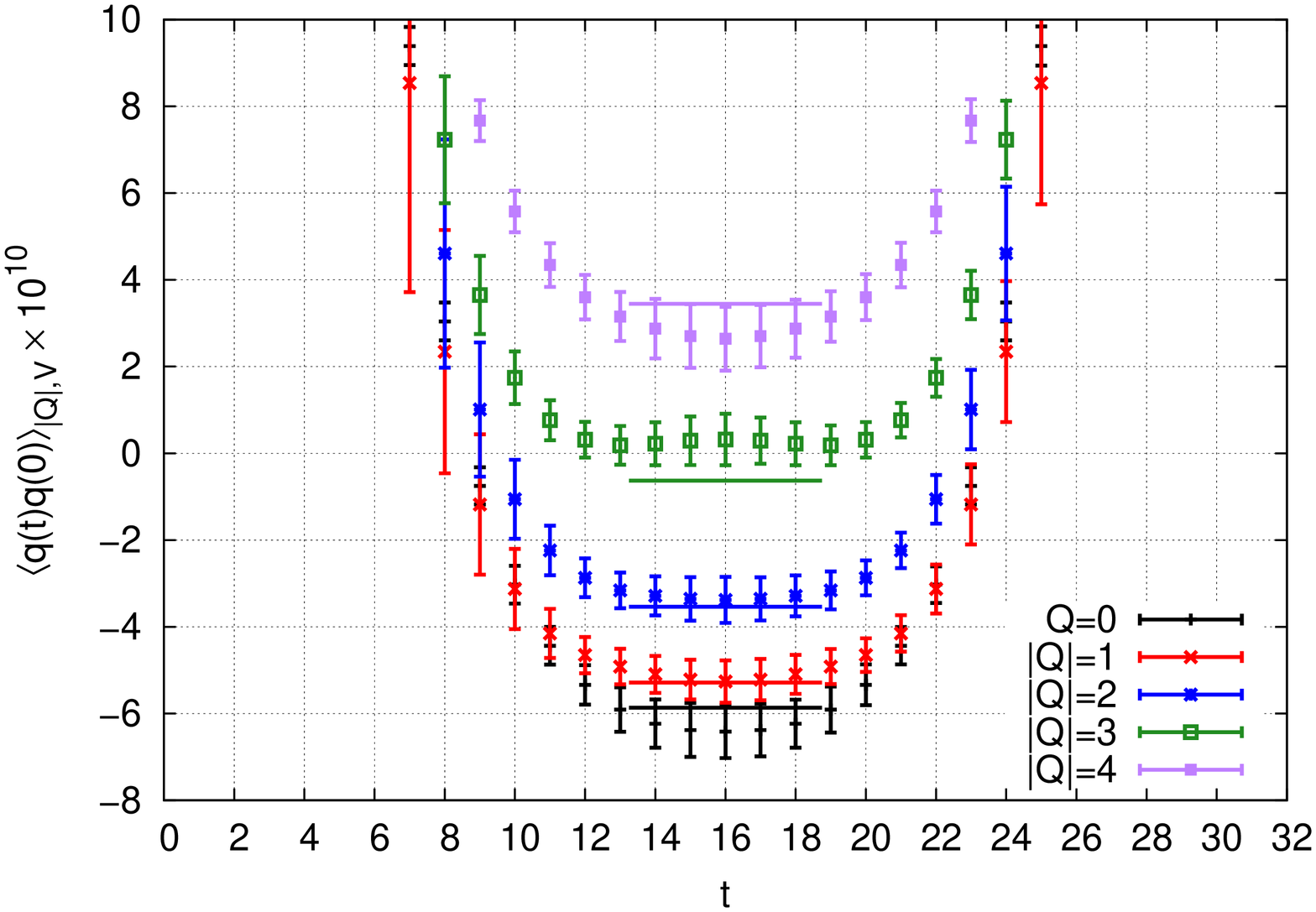}
\end{center}
\vspace*{-5mm}
\caption{Left: $\la q{'}^{\, 2} \ra$ in 2-flavour QCD, after 
$t = 5 t_{0}$, in $|Q| \leq 2$, and a global fit. Right: data and
fits for the AFHO method \protect\cite{AFHO}, 
cf.\ eq.\ (\protect\ref{AFHOeq}).}
\vspace*{-2mm}
\label{q2QCD}
\end{figure}
Fig.\ \ref{q2QCD} (left) 
shows data for $\la q{'}^{\, 2} \ra$ from
$|Q|= 0,\, 1,\, 2$,
after flow time $t = 5 t_{0}$ \cite{slabQCD}. At extreme values,
$x \gtapprox 0$ and $x \ltapprox 1$ (where thin slabs are involved),
the data deviate from a parabolic shape. This effect, caused by 
smoothing, is exponential; at $x \gtapprox 0$ we observed:
|deviation| $\propto \exp (-c(t) x)$.
Therefore we focus on the interval $0.2 \leq x \leq 0.8$, and
perform a joint fit --- of all data for $|Q| \leq 2$ ---
to the shifted parabola
\be  \label{q2add}
\la q{'}^{\, 2} \ra = V \chi_{\rm t} \, x (1-x) + {\rm const.} \ ,
\ee
which is shown in Fig.\ \ref{q2QCD}.
This fit works well, and it yields a result for $\chi_{\rm t}$,
which perfectly agrees with a direct measurement, and with the
result of the AFHO method \cite{AFHO} (cf.\ Section 1),
\be
\chi_{\rm t} \, a^{4} = \left\{ \begin{array}{ccc}
7.76(20) \cdot 10^{-5} & & {\rm direct} \\
7.63(14) \cdot 10^{-5} & & {\rm slab~method~for}~|Q| \leq 2 \\
7.69(22) \cdot 10^{-5} & & {\rm AFHO~method~for}~|Q| \leq 2 \ .
\end{array} \right.
\ee
Regarding the AFHO method, which refers to formula (\ref{AFHOeq}),
the correlations of the topological charge density and the plateau values
(after flow time $t = 6 t_{0}$) are shown in Fig.\ \ref{q2QCD} on the 
right.

\begin{figure}
\vspace*{-3mm}
\begin{center}
\includegraphics[angle=270,width=.45\linewidth]{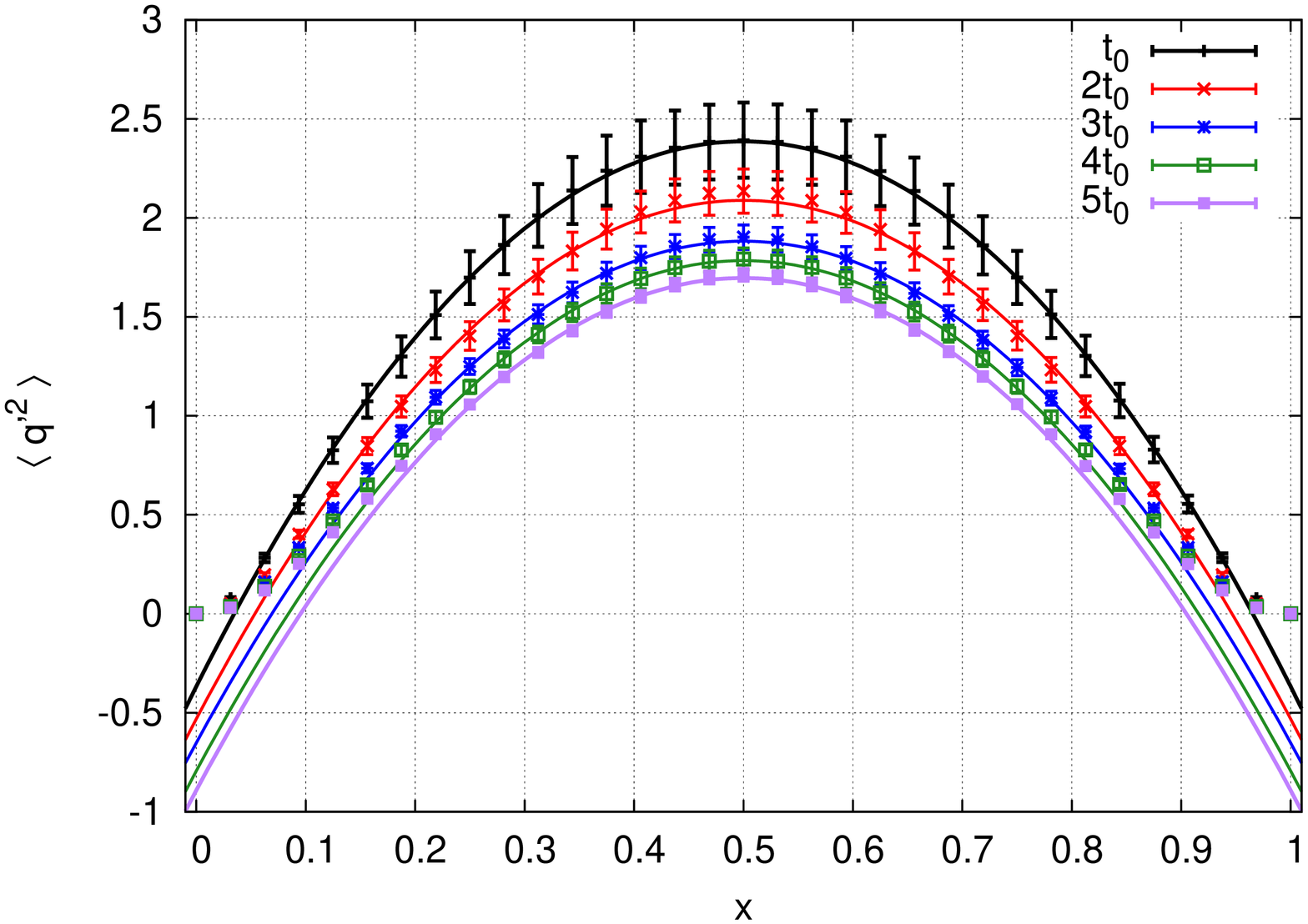}
\includegraphics[angle=270,width=.45\linewidth]{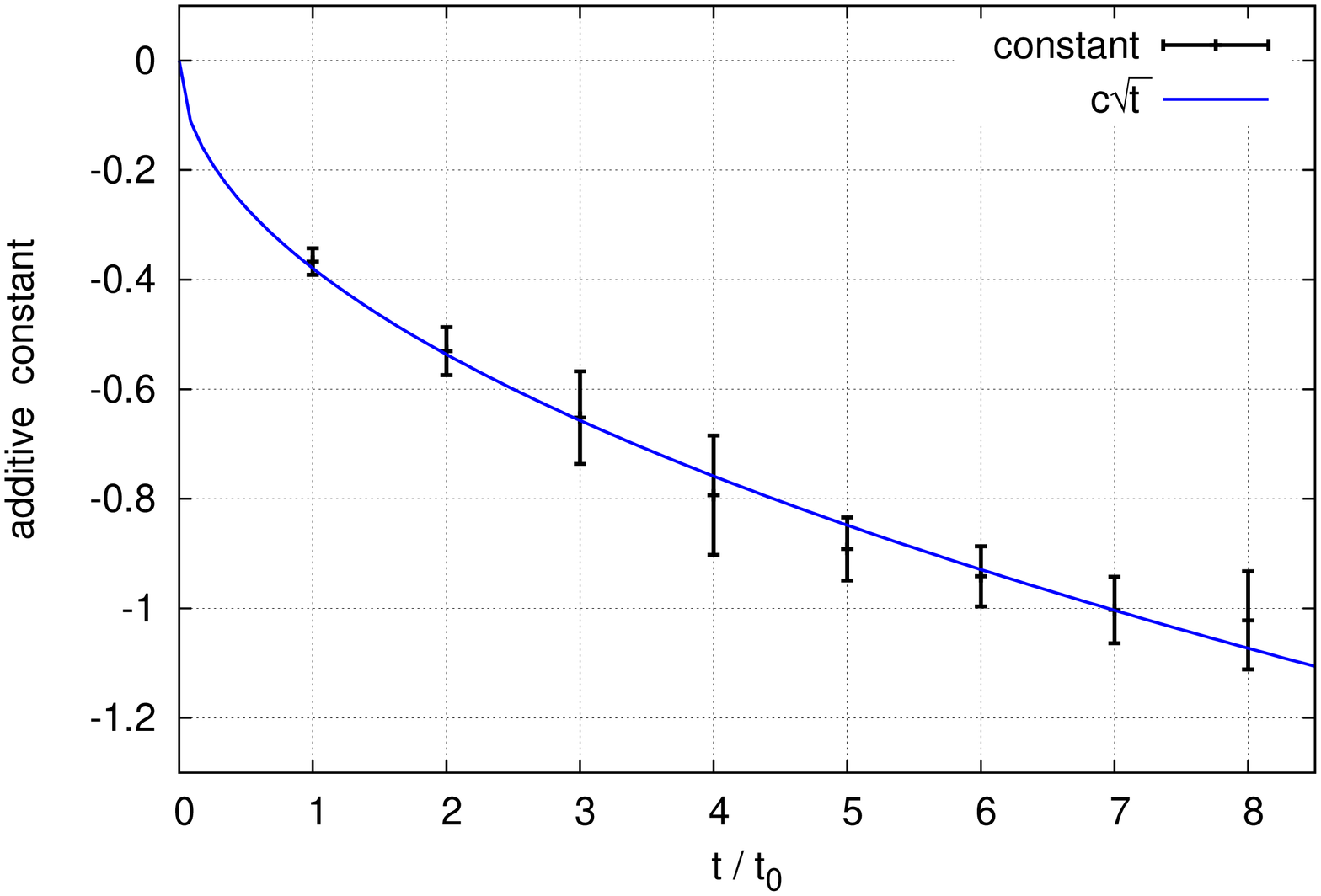}
\end{center}
\vspace*{-6mm}
\caption{Left: $\la q{'}^{\, 2} \ra$ in 2-flavour QCD at $|Q|=1$, 
at flow times $t = t_{0} \dots 5 t_{0}$. 
Even in the range $t= t_{0} \dots 8 t_{0}$,
the value for $\chi_{\rm t} \, a^{4} \cdot 10^{5}$ --- 
from a fit to eq.\ (\protect\ref{q2add}) --- is stable within errors 
({\it e.g.} $t_{0}\, :\, 7.70(20)$,
$2t_{0} \, : \, 7.69(21)$, $4t_{0} \, : \, 7.67(18)$, 
$6t_{0} \, : \, 7.80(18)$, $8t_{0} \, : \, 7.90(20)$).
Right: the additive const.\ of eq.\ (\protect\ref{q2add}) 
as a function of $t$.}
\label{flowtime}
\vspace*{-1mm}
\end{figure}

Fig.\ \ref{flowtime} (left) illustrates the evolution of 
$\la q{'}^{\, 2} \ra$ for flow time
$t_{0} \dots 5 t_{0}$, in the sector with $Q = |1|$ (as an example).
Longer flow time reduces the statistical errors (the
configurations are smoother), but the deviations at the extreme
values of $x$ are enhanced, and the additive constant in
eq.\ (\ref{q2add}) becomes more negative.
This constant is required here, but it has
not been anticipated in the slab formula (\ref{slabeq}). 
The plot in Fig.\ \ref{flowtime} on the right 
shows that it is consistent with a behaviour 
${\rm const.} \propto \sqrt{t}$, which corresponds to a diffusion
process. If we fit the data to the ansatz $c_{1} \sqrt{t}+c_{2}$,
we obtain $c_{2} = 0.003(18)$, which confirms that this constant
(practically) vanishes at $t=0$.

\section{Conclusions}
\vspace*{-3.9mm}

The slab method is a simple and robust procedure to measure
$\chi_{\rm t}$ within a fixed topological sector.
Hence it is not affected by ``topological slowing down'',
and it hardly costs any computing time, but there are 
persistent finite size effects (they tend to be polynomial at 
fixed topology \cite{BCNW,MarcArt,BCNWnum,AFHO,qqpap,slabpap}).
It works best at small $|Q|$, which is also the case for the
alternative fixed topology methods of Refs.\ \cite{BCNW,AFHO}.
In contrast to them, however, the {\em only} assumption needed 
for the slab method is a Gaussian distribution of the topological 
charges, which holds to a very good approximation.\footnote{A
generalisation which incorporates higher moments in the 
$Q$-distribution is feasible as well.}

We reviewed successful tests in O($N$) models \cite{slabpap} 
and in 2-flavour QCD \cite{slabQCD}. 
In the 2d O(3) model we obtained correct
results for $\chi_{\rm t}$ to $\%$-level, and in the 1d O(2) model
far beyond. In 2-flavour QCD, $\%$-level precision is attained 
as well, after gradient flow times $t = t_{0} \dots 8 t_{0}$. 
Here an additive constant is required in the fit, and one has to exclude
small intervals of $x$ close to 0 and 1.

\end{document}